\documentclass[prl,twocolumn,showpacs,preprintnumbers,amsmath,amssymb]{revtex4}
\usepackage{graphicx}% Include figure files
\usepackage{dcolumn}% Align table columns on decimal point
\usepackage{bm}% bold math

\begin{document}

\input{epsf}
\preprint{}

\title{Ferromagnetic phase transition and Bose-Einstein condensation in spinor 
Bose gases}

\author{Qiang Gu and Richard A. Klemm}

\affiliation{Max-Planck-Institut f\"{u}r Physik komplexer Systeme, 
N\"{o}thnitzer Stra{\ss}e 38, 01187 Dresden, Germany}

\date{\today}

\begin{abstract}
Phase transitions in spinor Bose gases with ferromagnetic (FM) couplings are 
studied via mean-field theory. We show that an infinitesimal value of the 
coupling can induce a FM phase transition at a finite temperature always above 
the critical temperature of Bose-Einstein condensation. This contrasts sharply 
with the case of Fermi gases, in which the Stoner coupling $I_s$ can not lead 
to a FM phase transition unless it is larger than a threshold value $I_0$. The 
FM coupling also increases the critical temperatures of both the ferromagnetic 
transition and the Bose-Einstein condensation.
\end{abstract}


\maketitle

The magnetism of Fermi (electron) gases has long been a research topic in solid 
state physics. Although many open questions remain, the magnetic properties of 
Fermi gases have been well understood\cite{mohn}. The Fermi surface plays an 
important role in determining their magnetism. For example, a magnetization 
density ${\overline M}$ in Fermi gases increases the band energy by splitting 
the Fermi surfaces for spin-up and spin-down particles. As a result, Fermi 
gases mainly behave as Pauli paramagnets in the absence of an exchange 
interaction. If an effective ferromagnetic (FM) exchange $I_s$ is present, 
electron gases can exhibit ferromagnetism. Within the framework of the Stoner 
theory, $I_s$ results in a negative molecular field energy when ${\overline M}$ 
is finite. If $I_s>I_0$, the Stoner threshold, the value of the molecular field 
energy becomes larger than that of the increase of the band energy induced by 
${\overline M}$, and then a FM ground state is energetically favored.

The magnetism of Bose gases has been less studied. But since the 
realization of Bose-Einstein condensation (BEC) in ultracold atomic 
gases\cite{cornell}, more and more attention has been attracted to this topic, 
because the constituent atoms, such as $^{87}$Rb, $^{23}$Na, and $^7$Li 
usually have (hyperfine) spin degrees of freedom and thus a magnetic moment 
$m$. $m$ can be large in atoms such as $^{52}$Cr\cite{weinstein}, for which 
$m=6\mu_B$, where $\mu_B$ is the Bohr magneton. Since 1998, atomic 
gases can be confined and cooled in purely optical traps in which their spin 
degrees of freedom remain active, and therefore investigating their magnetic 
properties becomes experimentally possible\cite{ketterle}. 

Ferromagnetism in spinor bosons without any spin-dependent interactions has 
already been theoretically studied. As opposed to the case of fermions, the 
ground states of spinor bosons are degenerate and a ferromagnetic state is 
among the ground states\cite{suto,yang}. Furthermore, the spinor Bose gas is 
rather apt to be magnetized by an external magnetic field even at a finite 
temperature $T$, as long as $T<T_c$, the BEC critical 
temperature\cite{yamada,caramico,Simkin,gu}. 

In ultracold atomic gases, a Heisenberg-like exchange interaction is usually 
present, arising from spin-flip scattering processes. For example, the 
effective two-body interaction in spin-$1$ atoms was given by\cite{ho}, 
\begin{eqnarray}
V = ~:\int d^3{\bf r}
     \left[ \frac {c_0}2 n^2({\bf r}) 
   + \frac {c_s}2 {\bf S}({\bf r})\cdot {\bf S}({\bf r}) \right]: ,
\end{eqnarray}
where $n({\bf r})=\sum_{\sigma}\psi_{\sigma}^\dag ({\bf r}) 
\psi_{\sigma}({\bf r})$, 
$S^\alpha ({\bf r}) = \sum_{\sigma\sigma^{\prime}} 
\psi_{\sigma}^\dag ({\bf r}) S^\alpha_{\sigma\sigma^{\prime}} 
\psi_{\sigma^{\prime}}({\bf r})$ with $S^\alpha$ 
($\alpha=x,y,z$) being $3\times 3$ spin matrices, $\psi_\sigma({\bf r})$ is a 
field annihilation operator for an 
atom in the hyperfine state $|F=1,\sigma\rangle$ at point ${\bf r}$, and 
$\sigma=1,0,-1$. The normal ordering removes the spin-spin interaction of a 
particle with itself. In the special case of $^{87}$Rb, the exchange is 
ferromagnetic ($c_s<0$)\cite{burke,pu}.

Ferromagnetic couplings can also be generated by ``superexchange" processes in 
correlated spinor boson systems described by the boson Hubbard model. Recently, 
this model was used to describe Bose gases moving in a periodic optical 
lattice\cite{greiner}, and a lot of work has been done to discuss its 
properties\cite{fisher}. It has been shown that the spinor boson Hubbard model 
can be reduced to the {\sl ferromagnetic} Heisenberg model, provided the 
on-site repulsion is large enough\cite{yang},
\begin{eqnarray}
H = - I_H \sum_{{\langle ij \rangle}}{\bf S}_i\cdot {\bf S}_j ,
\end{eqnarray}
where $I_H$ is the Heisenberg exchange constant. 

The presence of FM couplings enriches the properties 
of spinor bosons significantly.  One can naturally imagine that a FM phase 
transition could be induced, at a temperature plausibly depending on the 
energy scale of the coupling. Thus, an interesting question arises as to how 
the FM transition interplays with the BEC. We focus on this question in this 
paper. Without loss of generality, we study a {\sl homogeneous} spinor Bose gas 
before considering trapped gases. Our theoretical results indicate that the 
FM phase transition takes place at a finite temperature $T_F$ {\sl always 
above $T_c$}, regardless of the size of the coupling. This contrasts sharply 
with the case of fermions. Figure 1 shows schematically the 
relation between $T_F$ and $I_s$ for both Bose and Fermi gases. More recently, 
Isoshima {\it et al.}\cite{isoshima} and Huang {\it et al.}\cite{huang2} 
studied BEC in trapped $F=1$ spinor bosons with FM couplings, but the FM 
transition was not considered.

\begin{figure}
\center{\epsfxsize=65mm \epsfysize=40mm \epsfbox{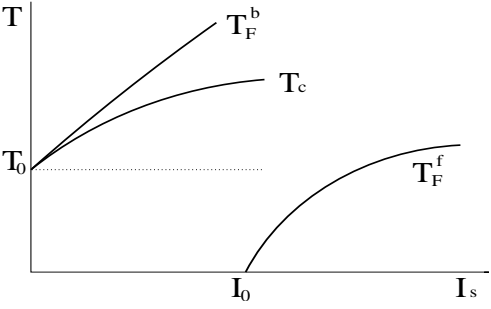}}
\caption{
\label{fig:epsart}
Schematic relations between transition temperatures and FM couplings $I_s$. 
$T^b_F$ and $T^f_F$ represent the FM transition temperature for Bose and Fermi 
gases respectively. $T_c$ and $T_0$ denote the BEC critical temperature for 
spinor bosons with and without couplings respectively. $I_0$ is the Stoner 
threshold.}
\end{figure}

{\sl The model.}--
Following the Stoner theory for fermion gases\cite{mohn}, a molecular field 
$H_m = I_s {\overline M}$ is introduced to describe the effective FM coupling 
between bosons, and the energy shift arising from the molecular field is 
\begin{eqnarray}
\epsilon^m_i = - H_m S^z_i = - I_s {\overline M} S^z_i .
\end{eqnarray}
The molecular field contains all the spin-dependent interactions. It can be 
derived from the exchange interactions via the mean-field approximation. 
In order to avoid the spin-spin interaction of a particle with itself, we treat 
the particles as being on some kind of ``lattice",
\begin{eqnarray}
-\sum_{\langle ij \rangle}{\bf S}_i\cdot {\bf S}_j &\approx& 
     - \sum_{\langle ij \rangle} ( \langle{\bf S}_i\rangle \cdot {\bf S}_j
     + {\bf S}_i \cdot \langle{\bf S}_j\rangle 
     -\langle{\bf S}_i\rangle \cdot \langle{\bf S}_j\rangle ) \nonumber\\
   &=&- Z {\overline M}\sum_{i} S^z_i + \frac 12 Z{\cal N} {\overline M}^2 ,
\end{eqnarray}
where $Z$ is the effective ``coordination number", which for a gas is an 
irrelevant dimensionless parameter of order unity, and ${\cal N}$ is the number 
of sites. Therefore $I_s=Z I_H$ for the Hamiltonian in Eq. (2). 
${\bf {\overline M}}_i=\langle{\bf S}_i\rangle$ serves as the ferromagnetic 
order parameter. We investigate a spinor Bose gas with hyperfine 
spin-$F$ and set the boson magnetic moment to unity. It is convenient to choose 
$\langle{\bf S}_i\rangle=(0,0,{\overline M}_i)$ and 
${\overline M}_i=\langle{S_i^z}\rangle 
= \sum_{\sigma}\sigma \langle \psi_{i\sigma}^\dag \psi_{i\sigma} \rangle$ where 
$\psi_{i\sigma}$ annihilates a boson with spin quantum number $\sigma$ at site 
$i$. We take ${\overline M}_i={\overline M}$ for a homogeneous boson gas.

Our main purpose is to show how the FM coupling affects the 
properties of spinor bosons. So we neglect spin-independent interactions for 
simplicity. Then the effective Hamiltonian for the grand canonical ensemble 
reads 
\begin{eqnarray}
H-N\mu= \sum_{{\bf k}\sigma} 
   [ \epsilon_{\bf k} - \mu - \sigma (H_m + H_e) ] n_{{\bf k}\sigma} ,
\end{eqnarray}
where $\epsilon_{\bf k}=\hbar^2k^2/2m^*$ is the kinetic energy of free 
particles with mass $m^*$, $\mu$ is the chemical potential, and $H_e$ is the 
external magnetic field. Since the Hamiltonian is diagonal, we may calculate the 
grand thermodynamical potential 
\begin{eqnarray}
\Omega &=& -\frac 1{\beta}{\rm ln} {\cal Z} 
  = -\frac 1{\beta} {\rm ln} {\rm Tr}{ e^{-\beta(H-N\mu)} } ,
\end{eqnarray}
where ${\cal Z} = {\rm Tr} \{ {\rm exp}[-\beta(H-N\mu)] \}$ is the partition 
function, and $\beta=1/(k_BT)$. The density of particles is 
\begin{eqnarray}
{\overline n} 
  = - \frac 1V \left( \frac {\partial\Omega}{\partial \mu} \right)_{T,V} 
  = \frac 1V\sum_{{\bf k}\sigma} 
    \langle n_{{\bf k}\sigma} \rangle , 
\end{eqnarray}
where $V$ is the volume of the system, $N$ is the total number of particles, 
and ${\overline n}=N/V$. ${\overline M}$ is determined self-consistently by
\begin{eqnarray}
{\overline M} 
    = -\frac 1V \left( \frac {\partial\Omega}{\partial H_e} \right)_{T,V}
    =  \frac 1V\sum_{{\bf k}\sigma} 
      \sigma \langle n_{{\bf k}\sigma} \rangle .
\end{eqnarray}
The external magnetic field $H_e$ is set to zero in the following calculations. 

We now consider only $F=1$ bosons. Eqs. (7) and (8) lead to the basic equations 
determining the phase diagram of the spin-$1$ boson system,
\begin{subequations}
\begin{eqnarray}
1 &=& n_0 + t^{\frac 32} \left[ 
     f_{\frac 32}(a) + f_{\frac 32}(a+b) + f_{\frac 32}(a+2b)  \right] , \\
M &=& n_0 + 
   t^{\frac 32}\left[ f_{\frac 32}(a)  - f_{\frac 32}(a+2b) \right] ,
\end{eqnarray}
\end{subequations}
where $a=-(\mu+H_m)/(k_B T)$, $b=H_m/(k_B T)$, the reduced temperature and 
coupling are given by $t={k_B T m^*}/{( 2\pi\hbar^2{\overline n}^{2/3} )}$ 
and $I={I_s {\overline n}^{1/3} m^*}/{( 2\pi\hbar^2 )}$, respectively, 
${\overline n}_0$ is the condensate density, 
$n_0={\overline n}_0/{\overline n}$ is the condensate fraction, 
$M={\overline M}/{\overline n}$ is the normalized magnetization, and 
$f_{s}(a)$ is the polylogarithm function defined as\cite{robinson} 
\begin{eqnarray}
f_{s}(a) \equiv {\rm Li}_s( e^{-a}) 
         = \sum_{p=1}^{\infty}\frac {\left(e^{-a}\right)^p}{p^s} .
\end{eqnarray}
We note that $f_s(0)=\zeta(s)$, the Riemann zeta function. 

Eq. (9a) is the standard formula for the density of bosons, by which the 
reduced BEC critical temperature $t_c$ can be determined\cite{huang}. $a>0$ for 
$t>t_c$ and $a\to 0$ as $t\to t_c$ from above. ${\overline n}_0>0$ and $a=0$ 
for $t<t_c$. In deriving Eq. (9b), we assume that only the spin-$1$ component 
of the bosons can condense. This assumption is discussed in the following.

\begin{figure}
\center{\epsfxsize=65mm \epsfysize=50mm \epsfbox{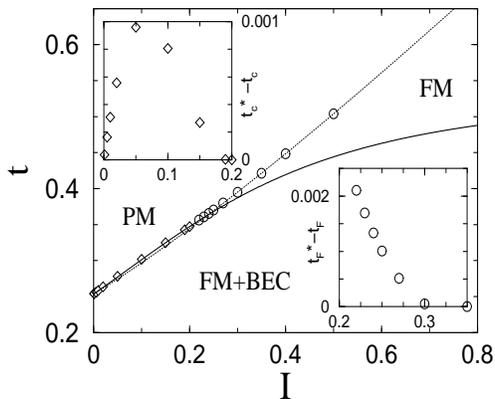}}
\caption{
\label{fig:epsart}
Reduced temperature $t$ vs. reduced FM coupling $I$ phase diagram of spin-1 
Bose gases. The paramagnetic normal phase (PM), ferromagnetic normal phase 
(FM), and coexisting phase of ferromagnetism and Bose condensation (FM+BEC) are 
indicated. The solid ($t_c$) and dotted ($t_F$) curves are results from 
Eqs. (15) and (11), respectively, and diamonds ($t_c^*$) and circles ($t_F^*$) 
are obtained from Eqs. (9). The  insets show the differences between $t_c^*$ 
and $t_c$, and between $t_F^*$ and $t_F$, respectively, versus $I$.}
\end{figure}

{\sl Preliminary analyses.}--
Assuming a FM phase transition is induced by $I$ below the reduced transition 
temperature $t_F$, it is of special interest to determine the relation between 
$t_F$ and $I$. We first suppose $I$ is very large so that $t_F>t_c$. Provided 
that the FM transition is {\sl continuous}, i.e., $b\to 0$ with $t \to t_F$, 
Eqs. (9) become
\begin{subequations}
\begin{eqnarray}
1 &=& 3 t_F^{\frac 32} f_{\frac 32}(a_F) ,\\
1 &=& 2 I t_F^{\frac 12} f_{\frac 12}(a_F) .
\end{eqnarray}
\end{subequations}
where $a_F=a(t_F)$. Equations (11) define a relation between $t_F$ and $I$. For 
a given $t_F$, $I$ is given by 
\begin{eqnarray}
I = \frac { \left[ 3f_{\frac 32}(a_F) \right]^{\frac 13} } 
    {2 f_{\frac 12}(a_F) } .
\end{eqnarray}
$I$ is a monotically decreasing function of $a_F$. As $a_F\to 0$, 
$f_{\frac 32}(a_F)\to \zeta(3/2)\approx 2.612$, and 
$f_{\frac 12}(a_F)\approx \sqrt{\pi/a_F}$\cite{robinson}. So for small values 
of $I$ and $a_F$ we have 
\begin{eqnarray}
a_F \approx {4\pi} t_0 I^2 ,
\end{eqnarray}
where $t_0=1/[3\zeta(3/2)]^{2/3}$ is the reduced BEC critical temperature for 
the Bose gas with $I=0$. Since $a_F$ is always larger than zero as long as $I$ 
is finite, Eq. (13) implies that {\sl an infinitesimal FM coupling can induce a 
FM phase transition at a finite temperature above the BEC critical 
temperature}, because $a_F$ is finite only for $t>t_c$. 

We now calculate the asymptotic behavior of the relation between $t_F$ and 
$I$ for small couplings: $I\ll 1$. From Eqs. (11) we find\cite{robinson}
\begin{eqnarray}
\frac {t_F}{t_0} \approx  1 + 8\pi t^2_0 I  .
\end{eqnarray}
This equation shows that {\sl the FM transition temperature increases with 
the FM coupling}. 

By decreasing $t$ further, the Bose gas will undergo BEC. Due to the molecular 
field, the spin-1 bosons have the lowest free energy. So the condensate 
contains only the spin-1 component and the other two boson spin components 
remain in their excited states. Since ${\overline n}_0=0$ at the BEC critical 
temperature, the self-consistent Eqs. (9) reduce to
\begin{subequations}
\begin{eqnarray}
1 &=& t_c^{\frac 32} \left[ \zeta(\frac 32)
   + f_{\frac 32}(b_c) + f_{\frac 32}(2b_c) \right] , \\
\frac {b_c} I &=& 
   t_c^{\frac 12}\left[ \zeta(\frac 32) - f_{\frac 32}(2b_c) \right] ,
\end{eqnarray}
\end{subequations}
where $b_c=b(t_c)$. These equations define the relation between $t_c$ and $I$. 
Since the FM phase transition occurs above $t_c$, the system has a finite 
magnetization at $t_c$. When $I\ll 1$ and $b_c$ is very small, Eq. (15a) 
leads to 
\begin{eqnarray}
\frac {t_c}{t_0} \approx
    1 + \frac {4(1+\sqrt{2})\sqrt{\pi b_c}}{9\zeta(\frac32)} .
\end{eqnarray}
Equation (16) is similar to the expression of the BEC critical temperature in 
an external magnetic field\cite{Simkin}, but here $b_c\approx 8\pi t_0I^2$ is 
not the external field, but is a self-consistently determined quantity 
proportional to the spontaneous magnetization. 
Substituting $b_c$ into Eq. (16), we find
\begin{eqnarray}
\frac {t_c}{t_0} \approx 
    1 + \frac 83 (2+\sqrt{2})\pi t^2_0 I .
\end{eqnarray}
This equation indicates that {\sl the FM coupling also increases the BEC 
critical temperature}.

\begin{figure}
\center{\epsfxsize=65mm \epsfysize=45mm \epsfbox{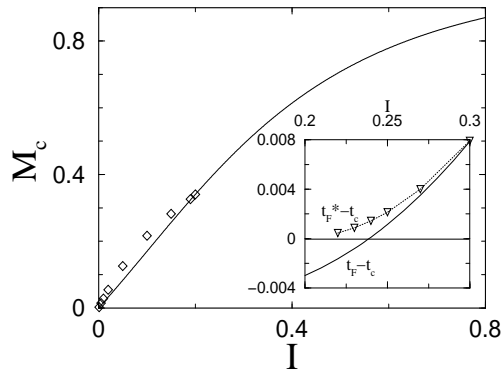}}
\caption{
\label{fig:epsart}
Plots of the normalized magnetization $M_c$ at reduced BEC critical temperature 
$t_c$ vs. reduced FM coupling $I$. The solid curve denotes $M_c$ at $t_c$ 
obtained from Eqs. (15), and the diamonds represent $M_c$ at $t_c^*$ obtained 
from Eqs. (9). Inset: plots of $t_F^*-t_c$ (inverted triangles), and $t_F-t_c$ 
(solid curve). The dotted curve is a guide to the eye.}
\end{figure}

\begin{figure}
\center{\epsfxsize=65mm \epsfysize=70mm \epsfbox{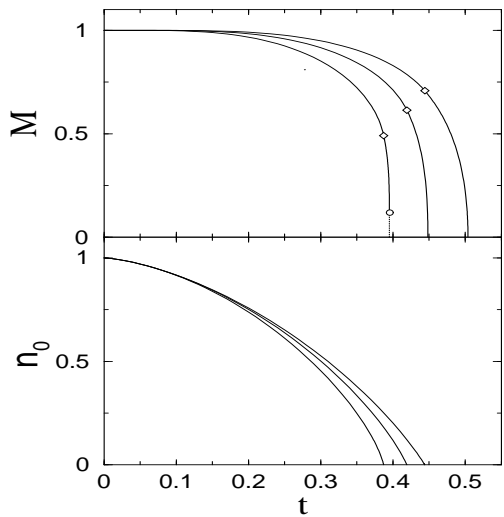}}
\caption{
\label{fig:epsart}
Order parameters of FM phase ($M$) and Bose condensed phase ($n_0$) versus the 
reduced temperature $t$ for the reduced FM couplings $I=0.5,0.4,0.3$, from top 
to bottom. The diamonds mark the points where BEC takes place. At $t_F^*$ 
(circle), $M$ drops discontinuously from $M_F=M(t_F^*)$ to zero.}
\end{figure}

{\sl Numerical results.}--
Comparing Eqs. (14) and (17), we see that $t_F$ is slightly smaller than $t_c$
for small FM couplings. This is unphysical and might be attributed to 
the assumption that the FM transition is continuous. To investigate this point, we 
solved the self-consistent Eqs. (9) numerically. The results obtained 
indicate that both the FM transition and the BEC become discontinuous if the FM 
coupling is sufficiently small. 

Fig. 2 shows the phase diagram of spin-1 Bose gases. 
We first discuss the Bose-Einstein condensation. For $I\gtrsim I_c\approx 0.2$, 
$t_c=t_c^*$,  as shown in the left inset, and both $a$ and ${\overline n}_0$ 
are exactly determined to be 
zero at $t_c$, while $M$ has a finite value: $M(t_c)=M_c$. In this case, the 
results from Eqs. (15) agree perfectly with those obtained from Eqs. (9). With 
increasing $I$, $t_c$($t^*_c$) tends to the upper limit value 
$1/[\zeta(3/2)]^{2/3}$ which is the reduced BEC critical temperature of scalar 
bosons. Nevertheless, when $I<I_c$, the BEC becomes a discontinuous phase 
transition: ${\overline n}_0$ has a small but nonzero value at $t^*_c$. In this 
case, $t^*_c$ is slightly larger than $t_c$, as the left inset shows. The fact 
that the Bose gas has a finite magnetization at 
$t_c$($t^*_c$) implies that a FM transition has taken place at a higher 
temperature as stated above. Fig. 3 shows the normalized magnetization $M_c$ at 
$t_c$($t^*_c$) versus the reduced FM coupling $I$.

Fig. 4 plots the order parameters of both phases. The normalized magnetization 
$M$ drops quickly with $t$ increasing above $t_c$($t^*_c$), and tends to zero 
at the reduced FM transition temperature $t^*_F$. $t^*_F=t_F$ when 
$I \gtrsim I_F\approx 0.35$, as shown in the right inset of Fig. 2. The smaller 
$I$ is, the more quickly the magnetization drops. When $I < I_F$, $M$ does not 
disappear continuously but drops from a finite value $M_F=M(t^*_F)\neq 0$ to 
zero abruptly at $t^*_F$, with $t^*_F>t_F$ as shown in the right inset of 
Fig. 2. This signals that the FM transition becomes first order. 
$M_F\approx 0.11$ for $I=0.3$. In this case Eqs. (11) no longer hold. We 
determine $t^*_F$ for the coupling down to $I=0.22$ and the results show that 
$t^*_F>t_c$, although $t_F<t_c$ when $I\lesssim 0.24$, as shown in the inset of 
Fig. 3. Hence the FM transition occurs at a higher temperature than the BEC. 
However, for small $I$, $t^*_F$ is so close to $t^*_c$ that it is very 
difficult to solve Eqs. (9) for the FM normal phase. 

It is worth noting that the FM transition is continuous for large couplings. 
This point is consistent with the Weiss molecular field theory for classical 
particles, in which the FM transition is continuous\cite{mohn}, because for 
large couplings, the FM transition occurs at a relatively high temperature, 
when the Bose statistics reduces to Boltzmann statistics.

In conclusion, we studied the ferromagnetic phase transition and 
Bose-Einstein condensation in spinor Bose gases with ferromagnetic couplings 
via mean-field theory. We showed that the coupling, regardless of its 
magnitude, induces a ferromagnetic phase transition at a temperature always 
above the critical temperature of Bose-Einstein condensation. Moreover, the 
ferromagnetic coupling also increases the critical temperatures of both phase 
transitions.

R.A.K acknowledges partial support from the Max-Planck-Institut f\"ur Chemische 
Physik fester Stoffe.


\begin{references}

\bibitem{mohn} P. Mohn, {\sl Magnetism in the Solid State: an Introduction} 
(Springer- Verlag, Berlin, 2003)

\bibitem{cornell} M.H. Anderson {\it et al}., Science {\bf 269}, 198 (1995);
K.B. Davis {\it et al}., Phys. Rev. Lett. {\bf 75}, 3969 (1995);
C.C. Bradley {\it et al}., {\sl ibid}. {\bf 75}, 1687 (1995).

\bibitem{weinstein} J.D. Weinstein {\it et al}., Phys. Rev. A {\bf 57}, 
R3173 (1998).

\bibitem{ketterle} D.M. Stamper-Kurn {\it et al}., Phys. Rev. Lett. {\bf 80}, 
2027 (1998); J. Stenger {\it et al}., Nature {\bf 396}, 345 (1998).

\bibitem{suto} A. S\"{u}t\H{o}, J. Phys. A {\bf 26}, 4689 (1993): 
E. Eisenberg and E.H. Lieb, Phys. Rev. Lett. {\bf 89}, 220403 (2002).

\bibitem{yang} K. Yang and Y.-Q. Li, Int. J. Mod. Phys. B {\bf 17}, 1027 (2003).

\bibitem{yamada} K. Yamada, Prog. Theor. Phys. {\bf 67}, 443 (1982).
 
\bibitem{caramico} A. Caramico D'Auria, L. De Cesare, and I. Rabuffo, 
Physica A {\bf 225}, 363 (1996).

\bibitem{Simkin}M.V. Simkin and E.G.D. Cohen, Phys. Rev. A {\bf 59}, 1528 
(1999).

\bibitem{gu} Q. Gu, Phys. Rev. A (in production).

\bibitem{ho} T.L. Ho, Phys. Rev. Lett. {\bf 81}, 742 (1998); 
T. Ohmi and K. Machida, J. Phys. Soc. Jpn. {\bf 67}, 1822 (1998).

\bibitem{burke} J.P. Burke, Jr. and J.L. Bohn, Phys. Rev. A {\bf 59}, 
1303 (1999).

\bibitem{pu} H. Pu, W. Zhang, and P. Meystre, Phys. Rev. Lett. {\bf 87}, 
140405 (2001); W. Zhang {\it et al}., {\sl ibid}. {\bf 88}, 060401 (2002).

\bibitem{greiner} M. Greiner {\it et al}., Nature {\bf 415}, 39 (2002).

\bibitem{fisher} M.P.A. Fisher {\it et al}., Phys. Rev. B. {\bf 40}, 546 (1989); 
G.G. Batrouni {\it et al.}, Phys. Rev. Lett. {\bf 89}, 117203 (2002).

\bibitem{isoshima} T. Isoshima, T. Ohmi, and K. Machida, J. Phys. Soc. Jpn. 
{\bf 69}, 3864 (2000).

\bibitem{huang2} W.-J. Huang, S.-C. Gou, and Y.-C. Tsai, Phys. Rev. A {\bf 65}, 
063610 (2002).

\bibitem{robinson} J.E. Robinson, Phys. Rev. {\bf 83}, 678 (1951); M.H. Lee, 
Phys. Rev. E {\bf 56}, 3909 (1997).

\bibitem{huang} K. Huang, {\sl Statistical Physics} (John Wiley \& Sons, New 
York, 1987).

\end{references}
\end{document}